\documentstyle[amssymb,aps,twocolumn]{revtex}
\begin{document}
\author{E. A. Jagla and C. A. Balseiro}
\address{{\it Comisi\'on Nacional de Energ\'\i a At\'omica}\\
{\it Centro At\'omico Bariloche and Instituto Balseiro}\\
{\it 8400, S. C. de Bariloche, Argentina}}
\title{Vortex structure and resistive transitions in high-Tc superconductors}
\maketitle

\begin{abstract}
The nature of the resistive transition for a current applied parallel to the
magnetic field in high-Tc materials is investigated by numerical simulation
on the three dimensional Josephson junction array model. It is shown by
using finite size scaling that {\em for samples with disorder} the critical
temperature $T_p$ for the {\it c} axis resistivity corresponds to a
percolation phase transition of vortex lines perpendicularly to the applied
field. The value of $T_p$ is higher than the critical temperature for $%
j\perp H$, but decreases with the thickness of the sample and with
anisotropy. We predict that critical behavior around $T_p$ should reflect in
experimentally accessible quantities, as the $I-V$ curves.
\end{abstract}


\section{Introduction}

Dissipation in high-Tc superconductors caused by motion of vortex lines has
become one of the richest fields in the phenomenology of these new materials.%
\cite{blatter} From the point of view of applications this dissipation is
one of the principal limiting factors when trying to achieve high critical
currents. From a purely theoretical point of view, due to the high critical
temperatures and to the layered structure, the physics of the vortex
structure becomes qualitatively different to that in the low-Tc materials.
The magnetic field-temperature ($H-T$) phase diagram is not well described
by a mean field approximation. An estimation of the fluctuation-dominated
region around the mean field phase transitions in the $H-T$ diagram shows
that this is quite accessible experimentally.

One of the new phases of the system that has been observed in the High Tc's
is the so-called liquid phase. This phase is characterized by the fact that
the vortex lines are easily movable by applying an external force on them by
putting a current $I$, i.e., the resistance of the system, defined as $%
R=\lim_{I\rightarrow 0}V/I$ is different from zero. This liquid phase is
found for high enough temperatures, whereas for $T$ lower than a well
defined value $T_i$, the resistance becomes zero. $T_i$ is the
superconducting transition temperature in the presence of the magnetic
field. The nature of the transition at $T_i$ depends on the disorder present
on the system. When the amount of disorder is very small, the low
temperature phase is an Abrikosov solid phase and the transition at $T_i$ is
a first order melting transition, with the resistivity having a jump at $T_i$%
.\cite{1orden} For higher disorder a vortex glass phase has been proposed
for $T<T_i$.\cite{vglass,vglasss} In this case the transition at $T_i$ is
second order and the resistivity displays critical scaling for $T$ close to $%
T_i$. In some cases a crossover due to thermal depinning of vortices occurs.%
\cite{mcapas,il}

All these features correspond to the case when the magnetic field (which we
always supposed applied along the {\it c} axis of the material) and the
external current are perpendicular to each other. When the current is
parallel to the magnetic field a naive argument would say that the
dissipation is zero (up to the zero field transition temperature $T_c$)
because the Lorentz force on the vortices is zero. However, it has been
found experimentally that the transition temperature in this case -which we
will call $T_p$- is lower than $T_c$. This behavior is related to the
thermal fluctuations of the vortex lines.

The point of wether $T_p$ coincides or is higher than $T_i$ has been highly
controversial. It has been found experimentally that for twinned YBaCuO
samples $T_p$ turns out to be higher than $T_i.$\cite{exp1} In other words,
the superconducting coherence is lost in two steps when raising the
temperature: along the {\it ab} plane at $T_i$ and along the {\it c} axis a $%
T_p$. In addition, $T_p$ seems to approach $T_i$ when increasing the
thickness of the sample.\cite{exp2} Theoretically, there have been much
discussion about the possibility of the loose of coherence in two successive
steps, although the results are not conclusive. \cite{bosones,feig}
Numerical simulations give support to the picture of one \cite
{stroud,teitel2,carneiro} or two \cite{teitel1} transitions depending on the
model and the parameters used to describe the problem.

It has been realized that a non-zero dissipation for an arbitrary small
applied current parallel to the field -i.e., a non-zero resistivity- would
require the existence of arbitrary large loops parallel to the {\it ab}
planes. \cite{feig,teitel1} Arbitrary large isolated loops have infinite
energy, and thus the thermal energy which is necessary to create them is
infinite. However, the superposition of a large amount of small loops could
generate an arbitrarily large one at finite temperatures. The question
arises of wether the generation of these large loops is a thermal crossover
or a real phase transition occurring at temperature $T_p$.

In this paper we will try to clarify these points, by performing numerical
simulations on the three dimensional (3D) Josephson junction array (JJA)
model, which has been used previously to study thermodynamical as well as
transport properties of high Tc superconductors. \cite{teitel1,jb1} As the
whole problem is too broad, we will concentrate here on the properties of
samples with disorder and not too high anisotropies, leaving the study of
the dependence on the amount of disorder for another work.\cite{jbtbp} It
will be shown than in this case -in coincidence with experiments- the 3D JJA
model {\em for finite values of the thickness }has two different transition
temperatures $T_i$ and $T_p$ for the onset of the resistivity in the {\it ab}
plane and along the {\it c} axis, respectively. The transition at $T_i$ is
-in the case of our model and for the values of paramenters used- a thermal
depinning of vortex lines, but in a real materials corresponds to a second
order phase transition -the vortex glass transition, which has been
extensively studied before. The temperature $T_p$ corresponds to the onset
of the resistivity along the {\it c} axis. We will show \cite{jb2} that $T_p$
is the threshold value of a percolation phase transition perpendicularly to
the applied field. Moreover, $T_p$ is a well defined thermodynamical
temperature for any value of the thickness of the sample, as long as the
{\it ab} plane can be considered as infinite. The precise value of $T_p$
decreases when increasing the thickness, suggesting that the two transitions
collapse onto a single one for a bulk 3D sample.

In order to show these properties of the transition at $T_p$, we perform
finite size scaling on the three 3D JJA model. We find that the system
exhibits critical behavior around $T_p$, which reflects, in particular, in a
critical scaling of the $I-V$ curves. The values of the critical exponents
are different to those find for the vortex glass transition at $T_i$ and
agree with those found for the {\it c} axis $V-I$ curves of real materials.

The remainder of the paper is organized as follows. In the next section we
describe the 3D JJA model that we use for a detailed investigation of the
finite size scaling of the transition at $T_p$. In section III a simplified
model for the percolation transition is introduced. This model allows to
derive quantitative expressions for the finite size scaling that can be
checked against the numerical simulations, and we do that in section IV,
showing that the numerical data scale as expected for a percolation
transition. In section V we focus on the consequences of a percolation
transition on the experiments, and indicate that this consequences are
indeed observed. Finally, in section VI we summarize and conclude.

\section{Description of the model}

For the numerical simulations, we have worked on the isotropic three
dimensional-resistively shunted-Josephson junction array model \cite{mppp}
on a cubic lattice. This model has been extensively used to derive sensible
results for thermodynamical as well as transport properties of the vortex
structure in high-Tc superconductors. When looking at equilibrium properties
it reduces to the 3D uniformly frustrated XY model.

The 3D-JJA is conformed by a set of Josephson junctions connecting nearest
neighbor nodes of -in our case- a cubic lattice. Each junction consists of
an ideal Josephson junction of critical current $I_c$, shunted by a normal
resistance $R$, and a Johnson noise generator, which accounts for the
temperature. Thus, the current $i$ through each junction is given by:
\begin{equation}
i=I_c\sin \Delta \phi -\frac 1R\frac{\partial \Delta \phi }{\partial t}+\eta
^T(t)  \label{i}
\end{equation}
where $\Delta \phi $ is the difference between the superconducting phases on
the two nodes connected by the link. A vortex in a given plaquette of the
model is characterized by a value of $2\pi $ when following the variation of
the superconducting phase around this plaquette. An external current applied
between two opposite sides of the sample produces a Lorentz force on the
vortices of the system, and may give rise to dissipation, which is
characterized by a finite value of the potential difference $\Delta V$
between the two contacts. This potential difference is calculated in terms
of the phases of the system by using the Josephson equation
\begin{equation}
\Delta V\sim \left\langle \frac{\partial \left( \varphi _1-\varphi _2\right)
}{\partial t}\right\rangle
\end{equation}
$\varphi _1$ and $\varphi _2$ being the phases of the contacts, and $%
\left\langle ...\right\rangle $ indicating a time average.

The system of equations (\ref{i}) are complemented by the condition of
conservation of currents on each node. The equations are numerically
integrated in time \cite{metodo} and different magnitudes are calculated
after thermalization at each temperature.

The boundary conditions (BC) are taken open in the {\it ab} plane. However,
if open BC in the {\it c} axis are used, there will be a finite force on an
isolated vortex at finite temperature if the top and bottom ends of the
vortex are not aligned. The dissipation -that is non-zero even in the linear
regime- caused by this net force turns out to be independent of the
thickness of the sample, and in this sense, it is only a surface effect.\cite
{jb1} In order to eliminate this spurious surface effect it is crucial to
use BC for the {\it c} direction that assure that each vortex line leaving
the sample at a given point of the bottom plane re-enters at the same point
of the top plane. Strict periodic BC on the phases $\varphi $ have this
property, however we would obtain that the voltage difference between top
and bottom planes is identically zero.

We use boundary conditions in the {\it c} direction that reduce the surfaces
effects drastically -although {\em does not} eliminate them completely. We
proceed as follows: we numerically integrate the equations of motion for an
open system one time step, and find the value of the phases in the bottom
and top plane $\varphi _i^B$ and $\varphi _i^T$. Then the phases in this two
planes are modified to $\tilde{\varphi}_i^B=\left( \varphi _i^B+\varphi
_i^T\right) /2+\bar{\varphi}/2$ and $\tilde{\varphi}_i^T=\left( \varphi
_i^B+\varphi _i^T\right) /2-\bar{\varphi}/2$, with $\bar{\varphi}=\varphi
_i^B-\varphi _i^T$ being the mean difference between the phases of top and
bottom planes. At this stage the equations are integrated another time step.
This process guarantees that the vortex configuration is periodic along the
{\it c} axis, allowing at the same time to calculate the voltage difference
when we apply a current. The relation between {\it c} axis resistivity and
percolation described in Section IV would not be observable if open boundary
conditions were used.

Flux conservation implies that every flux line going into a unit cell of our
lattice also goes out of the cell. When two vortices go into the same
elemental cell we cannot tell which one of the two outgoing vortices
correspond to each one of the ongoing vortices. We interpret this situation
as the meeting of two vortex lines. In a real material this corresponds to
two vortex lines being at a distance lower than the core size of the vortex.
At high enough temperatures the vortex structure is heavily interconnected,
and we can follow a vortex line starting from one side of the sample and
arriving to the opposite side. In this case we will say that the vortex
structure has percolated perpendicularly to the applied field. Due to the
finite size of the systems used, and to the dynamical evolution, percolation
is not expected to occur at every time, but only at a given fraction of the
total time, which depends on temperature. We evaluate the probability $P$
that there exists a vortex line crossing the system from one side to the
opposite as a function of temperature. This probability $P$ will be one of
the most important variables in our analysis.

The ideal Abrikosov (triangular) vortex lattice is frustrated on the
subyacent square lattice. The thermodynamical properties of the uniformly
frustrated XY model on a square lattice have strong and non-monotonic
variations as a function of the field, due to commensurability effects.\cite
{tcdeh} This is a spurious effect for us, because we are interested in the
simulation of systems as close to the real samples as possible. In this work
we concentrate on the case of disordered samples, in which an Abrikosov -or
any ordered- lattice does not exist at all, even at low temperatures. In
real samples this may be due to the existence of twinned boundaries or point
defects, and the physics of the low temperature phase may be that of a
vortex glass or simply a frozen (disordered) set of flux lines.

In the numerical calculations we simulate the disorder by considering the 3D
JJA system with a random (non-correlated) distribution of critical currents
of the junctions. The energy of a vortex in a single loop with one Josephson
junction of critical current $I_c$ is proportional to $I_c$, so when taking
a random distribution of critical currents the links with lowest $I_c$ act
as pinning centers of vortices. If the disorder -i.e, the dispersion of $I_c$%
- is high enough, then the configuration of the ground state will be
disordered, and not related to the commensurability of the vortex lattice
parameter and the subyacent square lattice. In Fig. \ref{t0} we show a top
view of the vortex lattice at zero temperature, as obtained when slowly
cooling down the system in the presence of an external field of value $0.2$
in units of flux quanta per plaquette. Two different results after the
annealing process for the same distribution of disorder are shown. They
correspond to a system of $16\times 16\times 8$ sites, with a squared random
distribution of critical currents ranged between $0.2$ and $1.8$, which is
the same distribution that we use throughout the paper. As can be seen, the
vortex structure is not the corresponding to an ordered system, and moreover
the vortex lines are not necessarily straight. Using this disordered JJA
system we have checked we obtain a quite smooth behavior of the quantities
we calculate (for example the resistivities) as a function of the field. It
must be keep in mind that the situation for samples without defects,
especially in the case of a triangular lattice of Josephson junctions, is
different to the one described here, \cite{mingo1,huse} and will be
addressed in another work.\cite{jbtbp}

\section{Percolation transition and finite size scaling.}

The transition at $T_i$ of our model corresponds to a depinning of vortex
lines from their equilibrium positions due to thermal activation. This
reflects in the form of the resistivity vs temperature curves (current along
the {\it ab} plane) which show a typical thermally activated behavior.\cite
{jb1,mcapas} In addition, the process is well described by a single vortex
model, indicating that collective effects are of minor importance. \cite
{jb1,nota}

On the other hand, the transition at $T_p$ is a collective effect. As we
said before, for $T>T_p$ there exist some paths crossing the sample
perpendicularly to the applied field, that are free to move when a current
parallel to the field is applied. We will show now that these paths appear
at $T_p$ due to a percolation phase transition of vortex lines.

We start by introducing a simple model which will allow us to check the
proposed percolation phase transition by using standard finite size scaling.
Let us consider the bond percolation problem on a cubic structure of size $%
L_{ab}\times L_{ab}\times L_c$, but with one important modification: the
vertical bonds are supposed to be present with probability one, whereas the
horizontal bonds are present with probability $p$. This means that in all
cases (for all values of $p$) a rigid squared lattice of lines piercing the
sample along the {\it c} axis (representing the vortex lines at $T=0$) is
present. We will consider the value of $p$ as a function of temperature (see
below), and look for the probability $P$ that there exists a path connecting
two opposite sides of the sample perpendicularly to the rigid lines. This
model has a percolation phase transition that can be easily characterized in
terms of the percolation transition of the corresponding two dimensional
bond percolation problem on the square lattice. For example, if we look at
the percolation probability $P_{2D}$ on a two--dimensional sample of size $%
L_{ab}\times L_{ab}$, it satisfies (in the limit of large $L_{ab}$) a
scaling relation of the form:
\begin{equation}
P_{2D}\left( p_{2D}\right) =f\left( \left( p_{2D}-p_c\right) L_{ab}^{1/\nu
}\right)  \label{p2d}
\end{equation}
where $f$ is a universal function.\cite{stauf} The values of $p_c$ and $\nu $
are known to be $p_c=0.599$ and $\nu =1.33$. The exponent $\nu $ is the one
controlling the divergency of a correlation length $\xi $, i.e., $\xi \sim
\left( p_{2D}-p_c\right) ^{-\nu }$. This correlation length measures the
typical size of a cluster in the system.

The value of $P$ for our model can be read out from the solution for the
percolation probability of the two dimensional case, simply noting that two
neighbor rigid lines are connected if there is at least one horizontal
segment between them. This corresponds to replace $p_{2D}$ in Eq. (\ref{p2d}%
) by $1-\left( 1-p\right) ^{L_c}$. So, we obtain
\begin{equation}
P\left( p\right) =f\left( \tilde{x}L_{ab}^{1/\nu }\right)  \label{scale}
\end{equation}
with $\tilde{x}=1-p_c-\left( 1-p\right) ^{L_c}.$ Similar scaling expressions
for other quantities are readily derived.

We will tentatively apply these scaling expressions to the vortex structure.
Before showing the results, it is important to point out the differences
between the two situations: the percolation of a real vortex structure is a
complicated interacting percolation problem, which proceeds via thermal
fluctuations of vortex lines and thermally generated loops of different
sizes and energy scales. Moreover, only paths which conserve the direction
of the magnetic flux has to be considered. In the simplify model, instead,
we use a single probability $p$, which is related to the temperature by a
Boltzmann factor, i.e., $p\sim \exp \left( -\Delta /kT\right) /\left( 1+\exp
\left( -\Delta /kT\right) \right) $, with $\Delta $ being a energy scale
that is roughly given by the energy of a vortex loop connecting two nearest
vortices in the real system. In terms of the temperature, the scaling
variable $\tilde{x}$ reads
\begin{equation}
\tilde{x}=1-p_c-\left( 1+\exp \left( -\Delta /k_BT\right) \right) ^{-L_c}.
\label{variable}
\end{equation}

\section{Numerical finite size scaling}

We first show in Fig. \ref{pnoesc} the results for the percolation
probability $P$ obtained for the JJA model as a function of temperature -in
units of the mean Josephson energy of each junction- for an external
magnetic field $H=0.2$ flux quanta per plaquette, which is the value that we
used in all simulations. The results for different values of $L_{ab}$ and $%
L_c$, as well as the best fitting to equations (\ref{scale}) and (\ref
{variable}) are shown. The free parameters are $p_c$, $\Delta $, and $\nu $.
As can be seen in Fig. \ref{pnoesc}(c), the scaling is quite good in spite
of the approximations involved, giving confidence on the correctness of our
ideas. The values obtained for the parameters are $\nu =1.1\pm 0.1$, $%
p_c=0.5\pm 0.1$ and $\Delta =3.60\pm 0.05$. Although these parameters were
considered as free variables, the relation of $p_c$ and $\nu $ with the
values corresponding to the 2D model ($p_c=0.599$, $\nu =1.33$) is
noteworthy.

The size effects due to the finite values of $L_{ab}$ and $L_c$ have very
different effects that can be read out from the plots in Fig. \ref{pnoesc}
(a) and (b): when $L_{ab}$ increases for a given value of $L_c$, the
percolation transition as a function of temperature becomes steeper,
indicating that we have a well defined percolation transition in the limit $%
L_{ab}\rightarrow \infty $. This is the usual finite size effect of a
two-dimensional second order phase transition. Note that this occurs for a
fixed (finite) value of the thicknesses $L_c$, and that the transition
temperature $T_p$ -defined as the temperature at which there is a jump in
the percolation probability for $L_{ab}\rightarrow \infty $- is well
defined. The threshold value $T_p$ depends on the thickness $L_c$, as can be
seen in Fig. \ref{pnoesc}(b) $T_p$ decreases as $L_c$ increases. For $L_c\gg
1$ the form of this dependence can be inferred by putting $\tilde{x}=0$ in
Eq. (\ref{variable}):
\begin{equation}
k_BT_p\sim \Delta /\ln \left( L_c\right) .  \label{tp}
\end{equation}

This indicates a weak decrease of the transition temperature for thicker
samples. This is easy to understand: the thicker the sample, the probability
of having a vortex loop connecting two nearest vortices increases due to the
fact that there are more places for creating the loop, the mathematical
expression for this fact being contained in equation (\ref{variable}).

It should be keep in mind that the dependence of the transition temperature
on thickness {\em is not} the usual dependence of a pseudo-critical
temperature on the system size, because here we {\em do} have a sharp
percolation transition at any $L_c$ (as long as $L_{ab}\rightarrow \infty $%
), and in fact the transition is of a two dimensional character.

{}From the experimental point of view, it would be interesting to have a
relation between the percolation transition and some directly accessible
quantity. This quantity turns out to be the {\it c} axis resistivity of the
sample. The relation between percolation and {\it c} axis resistivity is
more clearly seen in the model in which the cosine interactions of the
Josephson junctions are replaced by Villain interactions. In this case, the
model can be exactly mapped onto a problem in which -after integrating out a
gaussian part, related to spin waves- the only degrees of freedom that
remain correspond to the positions of the vortex lines.\cite{villa}
Moreover, the voltage between two points $A$ and $B$ of the sample is due to
the vortex movement and can be calculated in the following way: each time a
vortex line crosses a path joining $A$ and $B$, the phase difference $%
\varphi _{AB}$ between points $A$ and $B$ changes in $\pm 2\pi $, the sign
depending on the sign of $\vec{v}\times \vec{H}$, where $\vec{H}$ represents
the direction of the magnetic field within the vortex, and $\vec{v}$ is its
velocity. The mean voltage between $A$ and $B$ is given by $\varphi _{AB}/t$%
, where $t$ is the time of measurement. (Nota: The argument is asintotically
correct only when $t\rightarrow \infty $, in other case the result may
depend on the chosen path joining $A$ and $B$).

The relation between percolation and {\it c} axis dissipation can now be
understood in the following way: if there are mobile vortex lines crossing
the sample perpendicularly to the {\it c} axis, any current along this axis
dissipates, i.e., percolation of mobile vortex lines is a sufficient
condition for dissipation. If there are no stationary vortex lines
percolationg perpendicularly to the {\it c} axis, still vortex loops can be
thermally generated in the sample. These loops can cut and reconnect and
also blow out and leave the sample. The movement of a vortex loop will
generate a net voltage only in the case in which the loop is blown out of
the sample, in any other case only fluctuations in the voltage are possible.
If a vortex loop grow to infinity, it implies that it generated during some
time interval a percolation path across the sample, so percolation is a
necessary condition for dissipation. Note that the argument is not true in
the case of using open boundary condition along the {\it c} direction,
because in this case a single vortex can give rise to dissipation performing
only slight displacements from its equilibrium position.\cite{jb1}

The relation between percolation and {\it c} axis resistivity was shown
before in Ref. \cite{jb2}. It was argued that the behavior of $\rho _c$ vs $%
T $ near the threshold is similar to that of $S/(L_{ab}^2L_c^3)$, where $S$
is the volume of the percolation cluster (The percolation cluster $S$ is
defines as the number of elemental segments of vortex lines which are linked
to (at least) one percolation path) . The argument that led to this
conclusion was a counting of the vortex lines that are involved in the
dissipation process, as well as the assumption of the validity of a viscous
motion description of the vortex lines. Here we put it in the language of
scaling near the transition temperature. In the same way as we derived
expression (\ref{scale}), we can write a scaled relation for the {\it c}
axis resistivity which reads
\begin{equation}
L_{ab}^{\tau _1}L_c^{\tau _2}\rho _c\left( T\right) =g\left( \tilde{x}%
L_{ab}^{1/\nu }\right)  \label{scale2}
\end{equation}
with $g$ a new universal function, and $\tau _1$ and $\tau _2$ being new
critical exponents (the value of $\nu $ in this equation should be the same
as in Eq. (\ref{scale}), if there is only one divergent length scale $\xi
\sim \left( T-T_p\right) ^{-\nu }$ close to $T_p$). The {\it c} axis
resistivity for different values of $L_{ab}$ as well as the scaling
according to Eq. (\ref{scale2}) are shown in Fig. \ref{rodelab} for a system
with $L_z=8$ (the current used to calculate the resistivity is always 1/100
of the critical current at zero temperature in the corresponding direction).
The values for the exponents are $\nu _2=0.40\pm 0.05$ and $\tau _2=1.6\pm
0.2$ (the value of $\tau _2$ is the one used to do the fitting of Fig. \ref
{rep}) . The critical behavior of the resistivity in the limit $%
L_{ab}\rightarrow \infty $ is given by
\begin{equation}
\rho \sim \left( T-T_p\right) ^\eta ,  \label{ro de t}
\end{equation}
where $\eta $ is the exponent that determines the asymptotic behavior of the
function $g$ for large values of its argument, i.e., $\lim_{x\rightarrow
\infty }g\left( x\right) \sim x^\eta $. In the limit of very large $L_{ab}$,
the dependence on this length should cancel out in Eq. (\ref{scale2}), and
we get $\eta =\nu \tau _1$. With the previously found value of $\nu $ and
the value of $\tau _1$ from Fig. \ref{rodelab}(b) we find $\eta =0.45\pm 0.1$%
. The asymptotic behavior of $\rho (T)$ is shown in Fig. \ref{rodelab}(a) as
a dotted line.

The close relation between {\it c} axis resistivity and percolation can be
also seen in the following way. By combining Eqs. (\ref{scale}) and (\ref
{scale2}) we obtain
\begin{equation}
L_{ab}^{\tau _1}L_c^{\tau _2}\rho _c\left( T\right) =g\left( f^{-1}\left(
P\right) \right)  \label{scale3}
\end{equation}
In fact, for all the sizes considered, this scaling relation is satisfied,
as shown in Fig. \ref{rep}. This figure also shows clearly that the {\it c}
axis resistivity is different from zero only if the percolation probability
is different from zero.

The previous scaling theory can be successfully generalized to account for
the possibility of (weak) anisotropy in the system. The only change is that
now the typical energy of an excitation $\Delta $ should be replaced by a
value which depends on anisotropy. We introduce anisotropy in our JJA system
by diminishing the {\it c} axis mean critical current of the junctions and
at the same time increasing the {\it c} axis elemental resistance by the
same factor $a$. In a mean field approach, the dependence of $\Delta $ on
the parameter $a$ is of the form $\Delta =\Delta _0/\sqrt{a}$, with $\Delta
_0$ the energy parameter for the isotropic system. Eq. (\ref{tp}) becomes
\begin{equation}
k_BT_p\sim \Delta _0/\sqrt{a}\ln \left( L_c\right) ,  \label{tp2}
\end{equation}
making clear that anisotropy gives a much stronger decrease of $T_p$ than
that obtained when varying $L_c$.

The validity of Eq. (\ref{tp2}) is limited at very large $L_c$ or very high
anisotropies, where $T_p$ may become close to the transition temperature $%
T_i $ along the {\it ab} direction (see below). For the range of sizes
studied, the {\it ab} plane resistivity when a current of value 0.01 is
applied is shown in Fig. \ref{rab}. In all cases $T_i\simeq 0.6$, which is
always lower than the corresponding value of $T_p$, i.e., we are in the
region where the previous scaling is expected to be valid.

\section{Physical consequences of the percolation transition}

The fact that the system has only one relevant length scale $\xi $ that
diverges at $T_p$, has some consequences that can be checked experimentally,
and in fact, be the key point in order to test the adequacy of the theory to
the experiments.

{}From the experimental point of view, the transition at $T_i$ is a vortex
glass to liquid transition or a depinning of individual vortex lines
depending on the strength of the disorder. \cite{il} In our simulations $T_i$
always corresponds to a thermal depinning.\cite{jb1} The value of $T_i$ is
rather thickness independent for the sizes of the sample studied (see Fig.
\ref{rab}, and also Ref. \cite{jb1}, Fig. 3(b)).

The dependence of the percolation temperature $T_p$ -and thus, of the
threshold value for the {\it c} axis resistivity- as a function of the
thickness of the sample and the anisotropy is given in Eq. (\ref{tp2}). As
we said above, all the results presented here correspond to the case where
the percolation temperature $T_p$ is greater than $T_i$, and in fact this is
a condition for the previous scaling expressions to be valid. This is
because, in order to have dissipation along the {\it c} direction we must
have a percolation path across the sample, but it is also necessary that
this path is free to move. As the path consists of segments that go through
the horizontal planes, this implies that the temperature should be greater
than the value $T_i$ for the thermal depinning of vortices in the planes.
This indicates that percolation is necessary although not sufficient to have
dissipation in the {\it c} direction. Our previous scaling expressions and
the relation between $\rho _c$ and the percolation probability are valid
only if the condition $T_p>T_i$ is satisfied.

{}From the point of view of the numerical simulations, it is very difficult to
make $T_p$ close to $T_i$ by increasing the thickness, due to its
logarithmic dependence on $L_c$, but it is possible by changing the
anisotropy $a$. However, when changing the anisotropy, a new feature
appears: if there is no correlated disorder in the planes, the vortex
configuration in the limit $a\rightarrow 0$ will be dominated by the
independent pinning on each planes, with the Josephson vortices between
planes -which in this limit have an energy $E\rightarrow 0$- forming an
entangled set of vortex lines. If, on the other hand, the disorder is zero,
when $T\rightarrow 0$ the vortex lines are still straight lines. These
different possibilities reflect on the relative values of $T_p$ and $T_i$.

In Fig. \ref{titp} we plot the values of $T_i$ and $T_p$ as a function of
the anisotropy for a $24\times 24\times 16$ sample{\em . }The value of $T_p$
is given from the onset in the {\it c} axis resistivity and from the
threshold of the percolation transition. The first point to be noted is the
finite value to which $T_i$ tends for very high anisotropies. This value is
the depinning temperature of a single plane. The two ways of determining $%
T_p $ give similar results as long as $T_p>T_i$, and this happens if the
anisotropy is lower than a critical value $a_{cr}$. For $a>a_{cr}$ the
threshold for the percolation transition moves to very low temperatures,
whereas the onset of the resistivity behaves smoothly. The form of the curve
$T_p$ vs $a$ for $a<a_{cr}$ follows the form $a^{-1/2}$ in agreement with
Eq. (\ref{tp2}).

The situation for $a>a_{cr}$ is not so clear. The fact that the vortex loops
percolates at very low temperatures indicates that vortex lines are
entangled even for $T\rightarrow 0$, due to the disorder within the planes.
The dissipation along the {\it c} direction for $T<T_i$ may be due to the
movement of vortex lines lying {\em entirely} between two neighbor planes.
In our simulations disorder is a fundamental ingredient in order to have $%
T_p<T_i$. In {\em ordered} samples $T_i$ and $T_p$ collapse onto a single
value for high anisotropies,\cite{jbu,teitelpc} because the vortices remain
straight for $T\rightarrow 0$. In this case the threshold of the percolation
transition and the onset of the {\it c} axis resistivity coincide for all
anisotropies.\cite{jbu}

Experiments performed in twinned YBaCuO samples show a decreasing of $T_p$
with thickness. This is qualitatively similar to the predictions of Eq. (\ref
{tp2}). For the more anisotropic BiSrCaCuO samples, $T_p$ is equal to $T_i$,
at least when looking at them by resistivity measurements,\cite{ejec1}
whereas susceptibility measurements suggest that $T_p<T_i$, but the value of
$\rho _c$ for $T_p<T<T_i$ is too small to be detected.\cite{ejec2} This may
be similar to the behavior in Fig. \ref{titp} for $a>a_{cr}$, although more
precise experimental as well as numerical work is needed in order to make
the point clearer.

In what follows we discuss some experiments that can be used to discriminate
between a crossover or a real phase transition occurring at $T_p$. A
percolation transition at $T_p$ should reflect in the fact that the $I-V$
curves scale onto two universal curves for $T$ higher or lower than $T_p$.
This kind of scaling has been a convincing proof of the vortex glass phase
transition at $T_i$.

In our picture of a percolation transition we expect -according to scaling
arguments near a second order phase transition-\cite{blatter,vglasss} that
the $I-V$ curves for different temperatures plotted as $V/I\left(
T-T_p\right) ^\eta $ vs $I/\left( T-T_p\right) ^\nu $ show a universal
behavior -i.e., independent of the particular value of $T$-, depending only
on wether $T$ is greater or lower than $T_p$. The exponents $\eta $ and $\nu
$ can be extracted from the results obtained in the previous sections. The
value at which $V/I\left( T-T_p\right) ^\eta $ tends for $I\rightarrow 0$
when $T>T_p$ should be independent of temperature, so $V/I\sim \left(
T-T_p\right) ^\eta $ for low currents. Thus we see that $\eta $ is the
critical exponent for the resistivity as defined in Eq. (\ref{ro de t}). We
now show that $\nu $ is the same exponent defined in section III. The value
of $I/\left( T-T_p\right) ^\nu $ for which the dissipation starts to be
appreciable when $T<T_p$ should again be temperature independent. This value
corresponds to the minimum current $I_{cr}$ that is necessary to blow out
the largest vortex loops perpendicular to the applied current that are
present in thermal equilibrium. If these largest loops have linear
dimensions $\sim \xi $, and thus an area $\sim \xi ^2$, the energy $E_I$ of
this loops in the presence of the current $I$ is roughly given by $E_I=a\xi
-bI\xi ^2$, with $a$ and $b$ numerical constants, and it has a maximum of
value $E_I^M=a^2/4bI$ for $\xi =a/2bI$. The energy $E_0$ when there is no
current is simply $E_0=a\xi $. The loop will blow out when $E_0>E_I^M$, and
we get $I_{cr}\xi =$cte. As the typical length $\xi $ scales as $\left(
T-T_p\right) ^{-\nu }$ we conclude that $I/\left( T-T_p\right) ^\nu $ is the
appropriate scaling combination, thus justifying the use of $\nu $ for the
exponent. It should be noticed that the same conclusion can be obtained
using dimensional analysis,\cite{blatter,vglasss} just keeping in mind that
in our case the relevant dimensionality of our diverging clusters is two,
instead of three.

If the proposed mechanism for the transition at $T_p$ in the 3D JJA model is
in fact the one occurring in experiments, the values obtained for the
exponents $\nu $ and $\eta $ in both cases should be similar. Preliminary
results in YBaCuO samples\cite{dppc1} indicate that this scaling is rather
good with values for the exponents that agree with ours. Note that the
values $\nu \simeq 1.1$ and $\eta \simeq 0.45$ that we found for the
exponents are certainly different than those obtained for the vortex-glass
transition along the {\it ab} plane ($\nu \sim 1.7$ and $\eta \sim 7$).

In addition to the experimental investigation of the scaling of the $I-V$
curves, it would be interesting to study the same scaling numerically in the
3D JJA model. However, this is a difficult task because of the size effects
due to the finite value of $L_{ab}$. We have done only two partial checks of
the scaling: we found that the current $I_{cr}$ for which the dissipation
starts to be appreciable scales as $I_{cr}\sim \left( T_p-T\right) ^\nu $,
with the value of $\nu $ in agreement with the value obtained {\em %
independently} before . In addition, right at $T_p$ and for intermediate
values of the current, the $V-I$ characteristics show a power law behavior $%
V/I\sim I^\alpha $ with $\alpha \simeq \eta /\nu $, as expected from the
scaling analysis.\cite{blatter}

The last point we will discuss in order to check our percolation theory
deals with the behavior of the in-plane resistivity $\rho _{ab}$ as a
function of the {\it c} axis resistivity $\rho _c$ for different values of
the external magnetic field. It turns out that the form of the $\rho _{ab}$
vs $\rho _c$ curves for different fields can be used to distinguish
experimentally between a second order phase transition at $T_p$ or a
crossover due to finite thickness of the samples.

If there were only a unique phase transition at $T_i$, and the effects seen
at $T_p$ were only due to the finite thickness of the sample, the situation
would be as follows: The resistivity would be controlled by an anisotropic
correlation volume of dimensions $\xi _{ab}$ and $\xi _c$. Close to $T_i$ we
would have $\xi _{ab}\sim \left( T-T_i\right) ^{-\nu }$ and $\xi _c\sim
\left( T-T_i\right) ^{-\nu }$ (notice that the numerical factor may be
different in both cases), and thus at $T_i$ we have $\xi _{ab}\rightarrow
\infty $ and $\xi _c\rightarrow \infty $ (we suppose that $L_{ab}$ can be
considered as infinite in the real samples). The resistivity along the {\it c%
} axis will be zero as long as $\xi _c\gtrsim L_c$, and due to the very thin
samples used, the temperature $T_p$ at which $\xi _c=L_c$, turns out to be
higher than $T_i$. If the relation $\xi _{ab}/\xi _c$ is field independent,
then the form of the curves $\rho _{ab}$ vs $\rho _c$ will not depend on
field at all. Otherwise, if $\xi _{ab}/\xi _c$ depends on the actual value
of the field, then $\rho _{ab}$ vs $\rho _c$ for different fields will not
have any constraining relation.

If the system has a percolation phase transition at $T_p$, at which a
correlation length $\xi $ diverges, the situation will be the following: the
curves $\rho _{ab}$ vs $\rho _c$ may depend in a complicated way on field,
except at the point at which $\rho _c$ becomes zero. At this point the value
of $\rho _{ab}$ should be independent of the field, the coincidence of $\rho
_{ab}(\rho _c=0)$ for different fields can be explained within our proposed
percolation transition scheme: the value of the {\it ab} plane resistivity
can be calculated using the fluctuation-dissipation theorem. For a single
vortex this dissipation is proportional to the typical lateral squared
displacement of vortices divided by the temperature: $\sim \Delta x^2/T$.
For the whole system the dissipation is proportional to $\Delta x^2n/T$,
with $n$ the vortex density, which is proportional to the magnetic field. So
we get $\rho _{ab}\sim \Delta x^2H/T$. At the temperature at which $\rho _c$
becomes zero, the displacement $\Delta x$ is roughly given by the distance
between vortices, and goes as $1/\sqrt{H}$, and we obtain that $\rho
_{ab}(\rho _c=0)$ is field independent.\cite{ust}

Experimental results indicate that the second possibility is the one that is
realized in high-Tc materials.\cite{dppc1} It is observed that the point at
which $\rho _c$ becomes zero corresponds to the same value of $\rho _{ab}$
irrespective of the field, whereas the part of the curves for $\rho _c\neq 0$
does depend on field. This favors our proposal of a real phase transition
occurring at $T_p$, against the possibility of a crossover due to the
matching of a correlation length with the thickness of the sample.

\section{Summary and conclusions}

In this paper we presented a description of the {\it c} axis-resistive
transition in the mixed state of high-Tc superconductors for the case of
samples where (due to disorder and for no so high values of anisotropy) the
superconducting coherence is lost along the {\it c} axis at a temperature $%
T_p${\em \ higher} than the corresponding to the {\it ab} plane. We showed
that the idea of a percolation phase transition perpendicularly to the
applied field as the mechanism driving the {\it c} axis to a linearly
dissipative (resistive) state is supported by finite size scaling in the 3D
JJA model as well as transport measurements in YBaCuO samples.

The transition is a well defined thermodynamic transition as long as $%
L_{ab}\rightarrow \infty $, irrespective of the value of the thickness $L_c$%
. However, the threshold value $T_p$ decreases logarithmically with $L_c$.
The analysis of the $\rho _{ab}$ vs $\rho _c$ curves give support to the
idea of a real phase transition at $T_p$, against the possibility of a
crossover due to finite thickness with only one single transition at $T_i$.
Also, preliminary results for the critical scaling of the $I-V$ curves are
consistent with a phase transition at $T_p$, the critical exponents being
close to those found in the numerical simulations, and certainly different
than those found for the vortex glass transition.

As pointed out before, the case corresponding to clean samples (that can be
numerically simulated using triangular lattices) gives results qualitatively
different to the ones described here. In particular, the interplay between
the transitions for the {\it c} axis and for the {\it ab} plane gives rise,
in some cases, to a single first order transition. If disorder is then put
into the system, the transition remains first order up to a critical value
of the disorder. If the disorder is increased further then two separate
transitions (with $T_p>T_i$ if the anisotropy is weak) are recovered.\cite
{jbtbp}

The case of very anisotropic samples, where possibly $T_p<T_i$, still needs
further experimental as well as theoretical clarification. Resistivity
measurements in BiSrCaCuO samples have not been able to detect a region
where $\rho _c\neq 0$, and at the same time $\rho _{ab}=0$. However this
could be due to the very high aspect ratio of the samples, which prevents an
accurate determination of $\rho _c$. On the other hand, our simulations in
the 3D JJA model suggest that the case $T_p<T_i$ is obtained {\it only }in
the case where the vortex lattice -due to the effects of disorder- has
percolated even for $T\rightarrow 0$. For samples without disorder
simulations suggest $T_p=T_i$ for very anisotropic samples.

\section{Acknowledgments}

We thank D. O. L\'{o}pez and F. de la Cruz for very fruitful discussions and
for sharing with us experimental results prior to publication.

E. A. J. is supported by CONICET, C. A. B. is partially supported by CONICET.

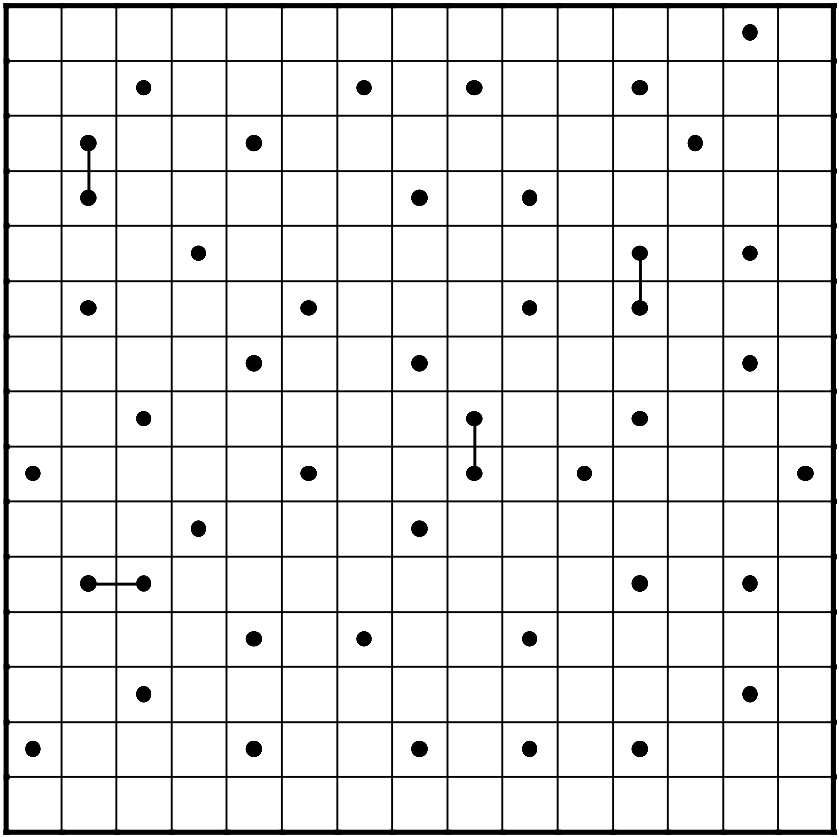
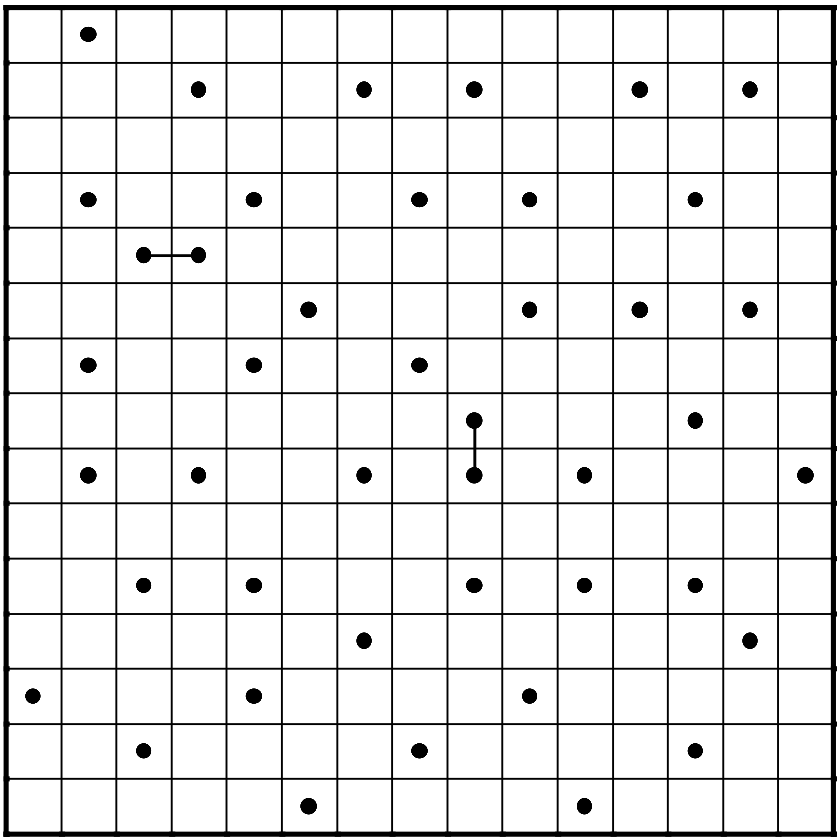
\begin{figure}[p]
\caption{Top view of the configuration of vortices in a lattice of $18\times
18\times 8$ for $H=0.2$ and a random distribution of disorder in the
critical currents of the junctions ($0.2<I_c<1.8$). Two different states
after an annealing process are shown. }
\label{t0}
\end{figure}
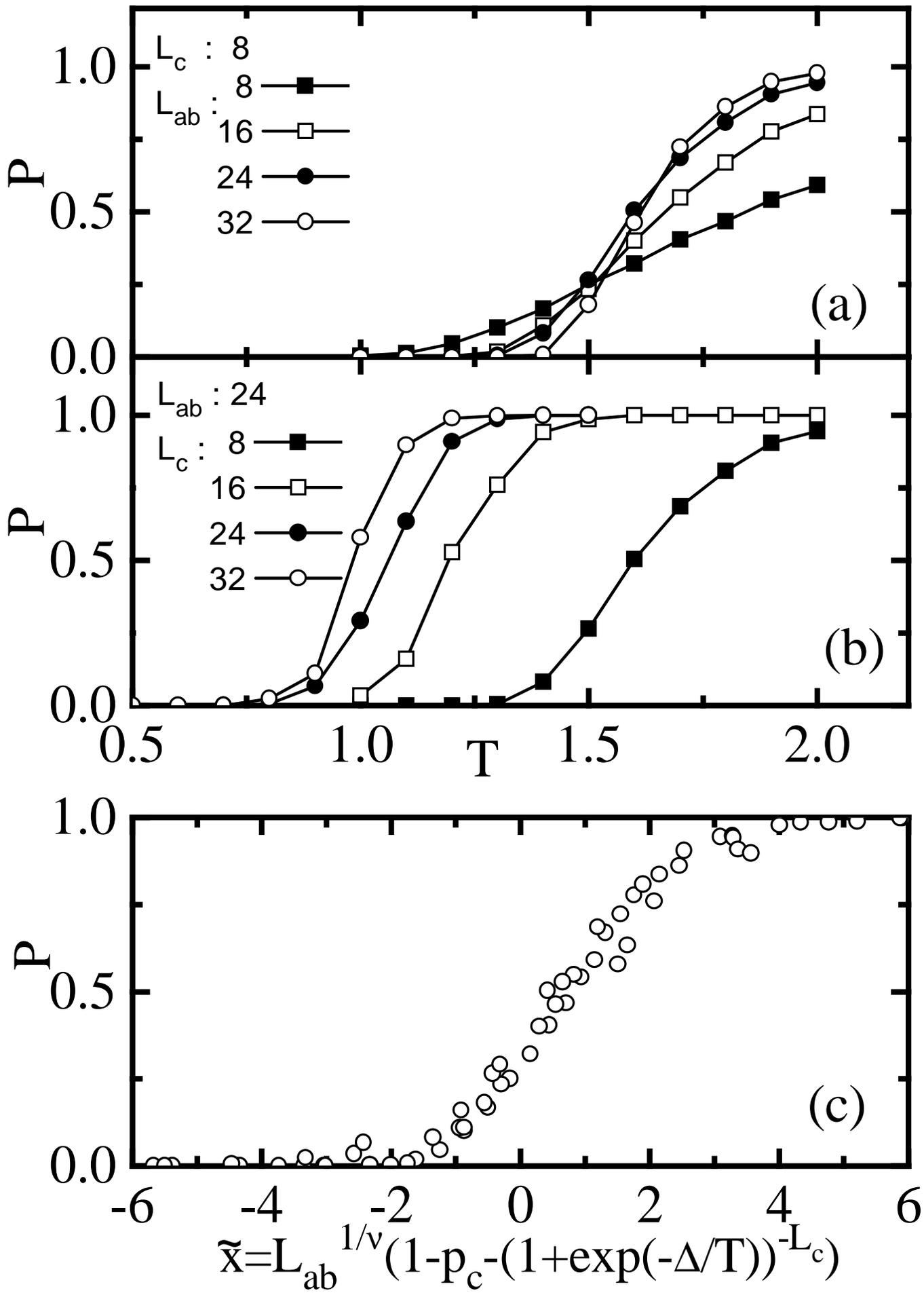
\begin{figure}[p]
\caption{Percolation probability for different values of $L_{ab}$ (a) and $%
L_c$ (b) as a function of temperature (in units of the mean Josephson energy
of the junctions). Panel (c): the curves from (a) and (b) scaled according
to the given formula (see text for explanation) with parameters $\nu =1.1$, $%
p_c=0.5$ and $\Delta =3.6$.}
\label{pnoesc}
\end{figure}
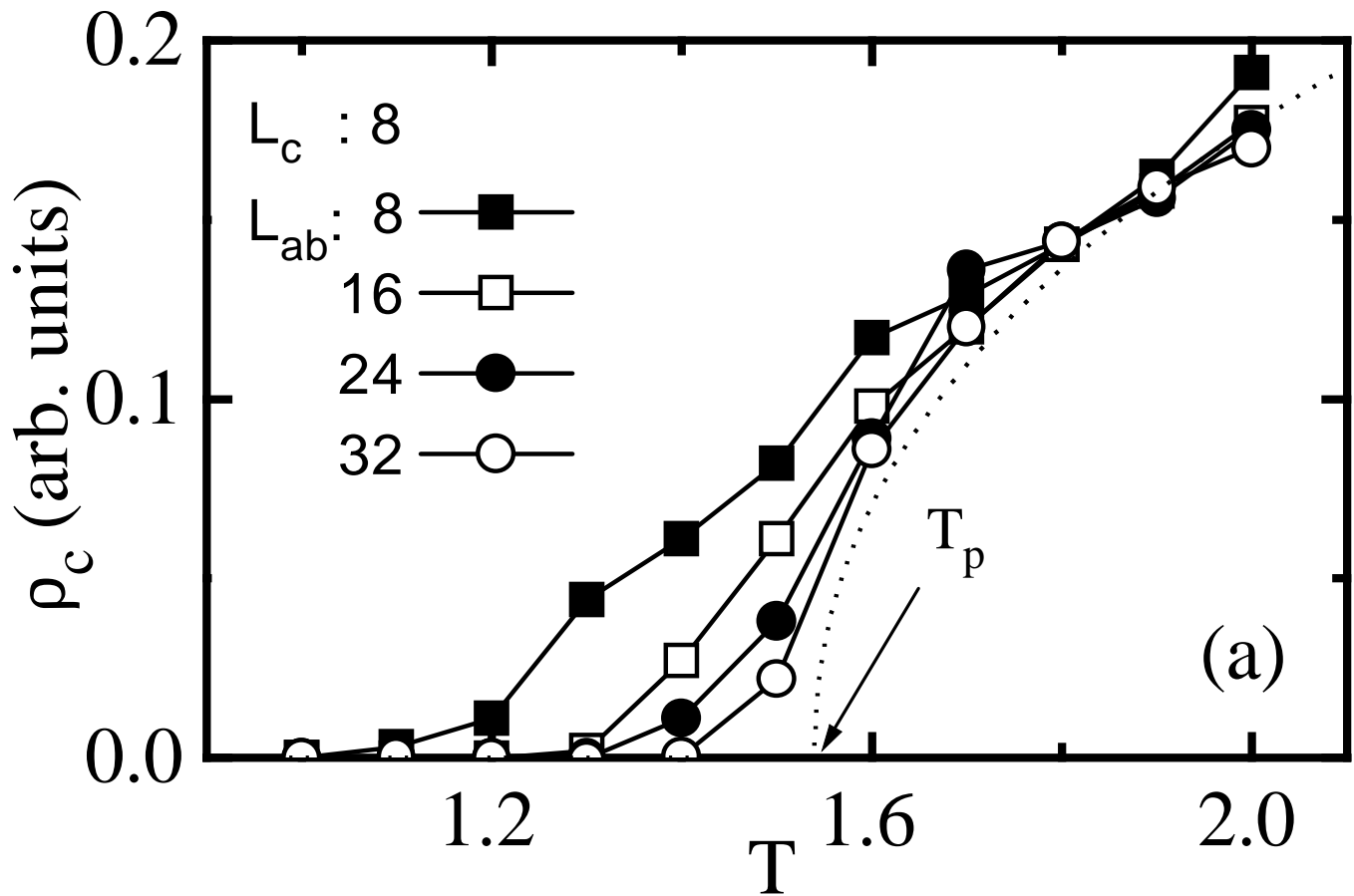
\begin{figure}[p]
\caption{(a) Scaled {\it c} axis resistivity as a function of temperature
for different values of $L_{ab}$, and (b) the same curves plotted against
the scaled variable $\tilde{x}$. The values of the exponents $\tau _1$ and $%
\tau _2$ are $\tau _1=0.4$ and $\tau _2=1.6$. In (a) the limit for large $%
L_{ab}$ -namely $\rho _c=\left( T-T_p\right) ^{\eta \tau _1}$- is also
shown. }
\label{rodelab}
\end{figure}
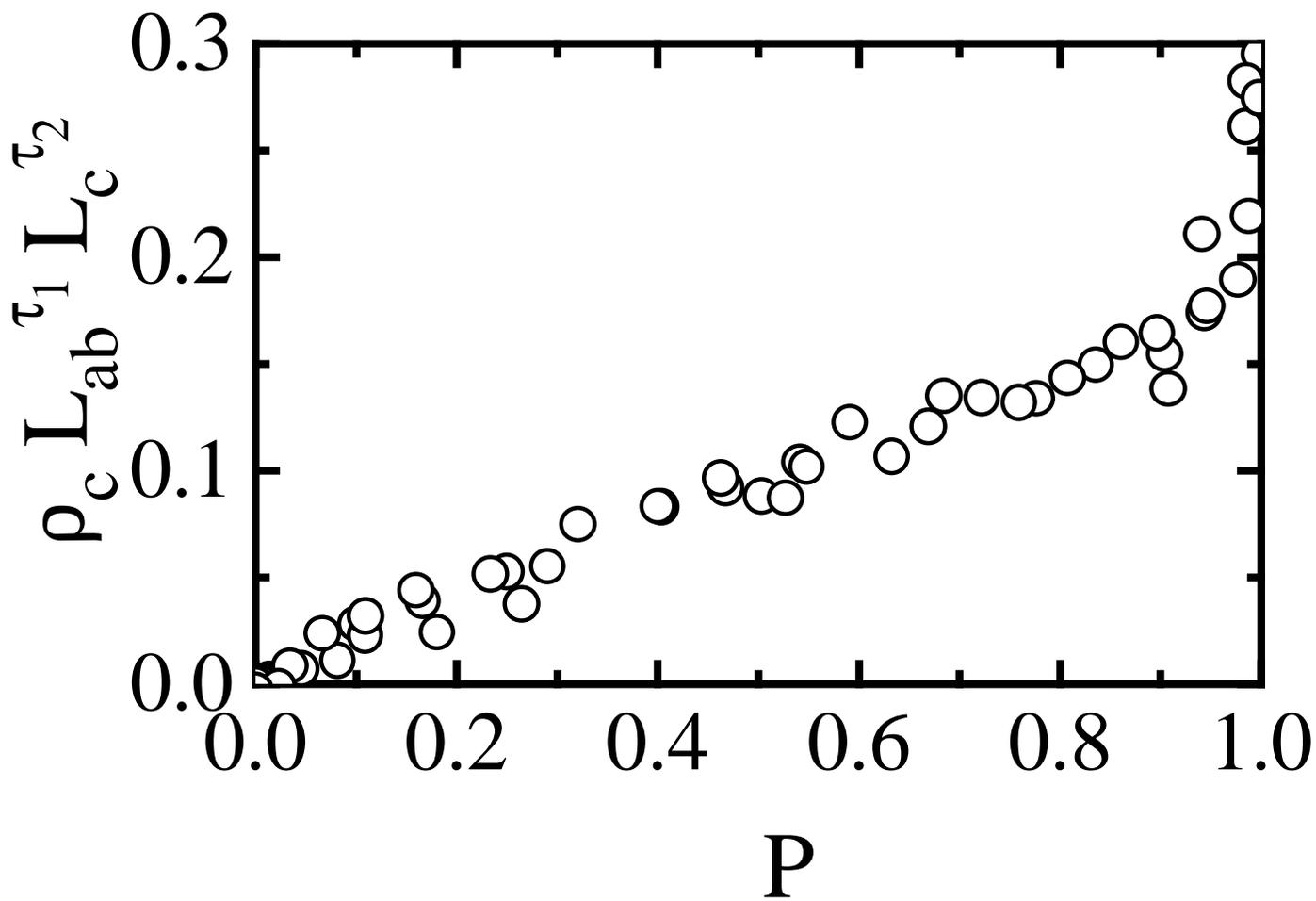
\begin{figure}[p]
\caption{Scaled {\it c} axis resistivity vs percolation probability for all
sizes quoted in Fig. 2(a) and (b).}
\label{rep}
\end{figure}
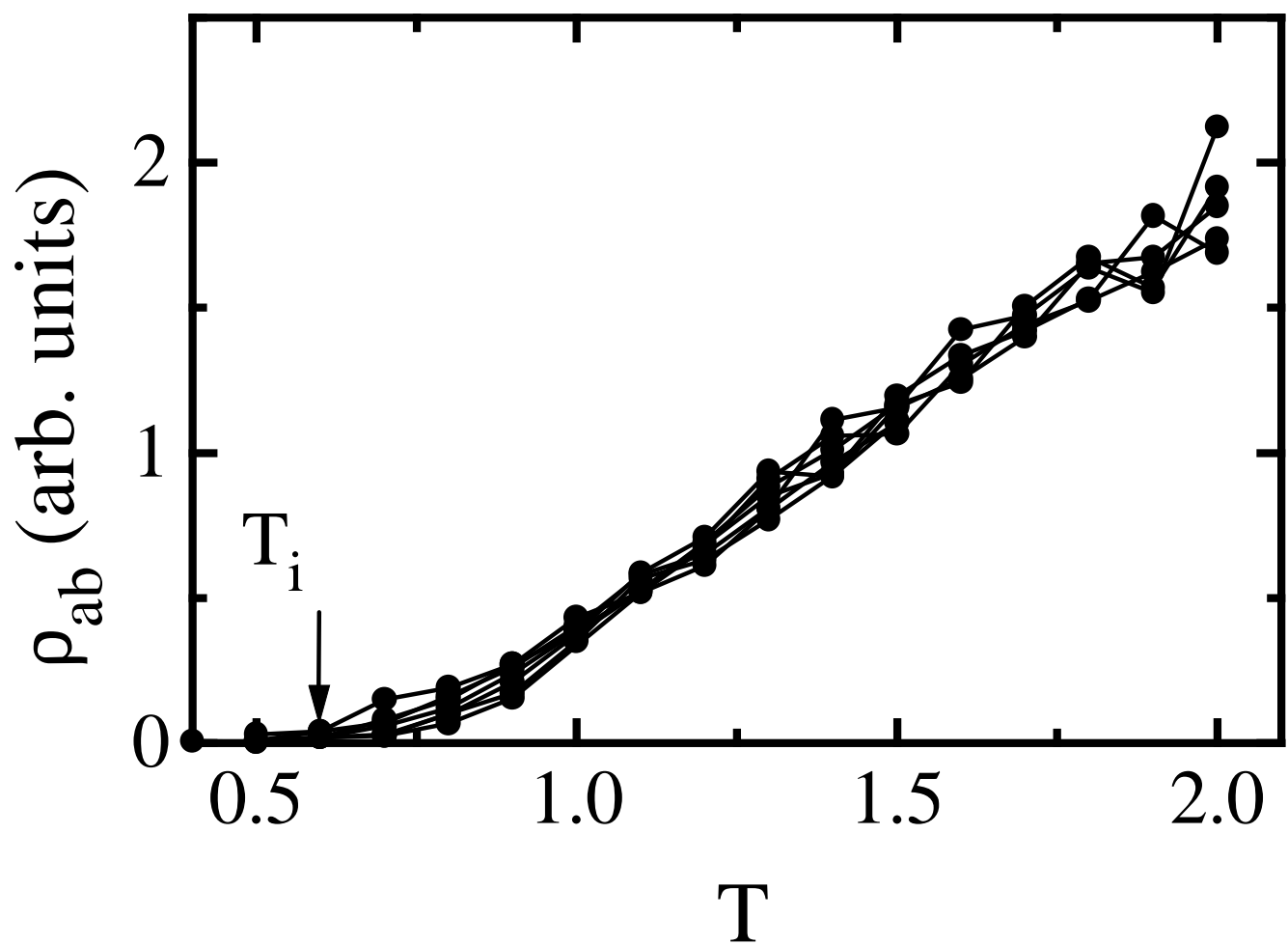
\begin{figure}[p]
\caption{{\it ab} plane resistivity for all system sizes considered (as
quoted in Fig. 2(a) and (b)). All curves have the critical temperature
around $T_i\simeq 0.6$.}
\label{rab}
\end{figure}
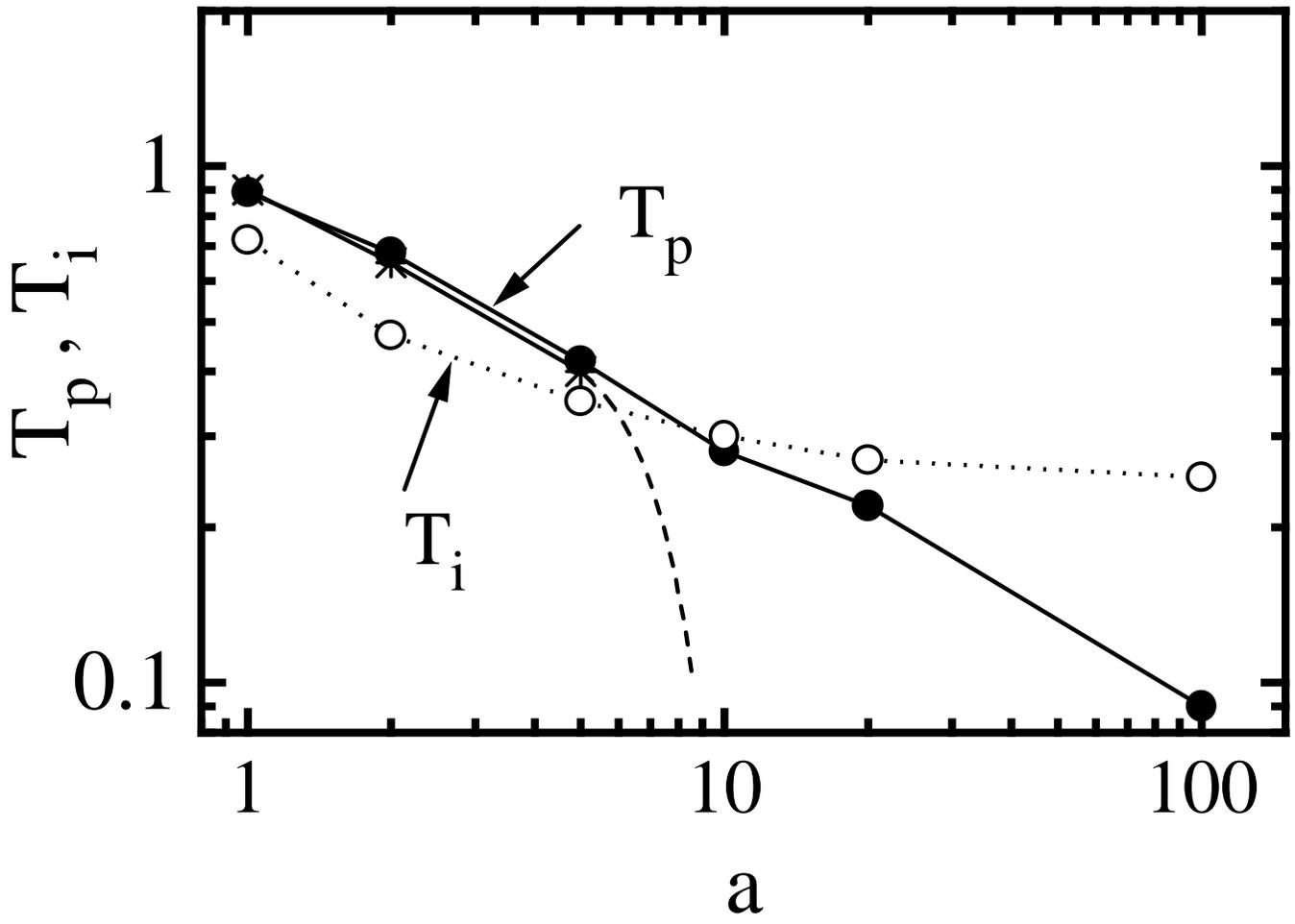
\begin{figure}[p]
\caption{Critical temperatures for the {\it c} axis ($T_p$) and for the {\it %
ab} plane ($T_i$) as a function of anisotropy for a $24\times 24\times 16$
sample. For $T_p$, circles correspond to the onset of the resistivity,
whereas stars indicate the onset of the percolation. For $a\geqslant 10$ the
onset of percolation moves towards $T=0$.}
\label{titp}
\end{figure}


\begin{references}
\bibitem{blatter}  G. Blatter, M. V. Feigelman, V. B. Geshkenbein, A. I.
Larkin, and V. M. Vinokur, Rev. Mod. Phys. {\bf 66}, 1125 (1994).

\bibitem{1orden}  H. Safar {\it et al}, Phys. Rev. Lett. {\bf 69}, 824
(1992).

\bibitem{vglass}  M. P. A. Fisher, Phis. Rev. Lett. {\bf 62}, 1415 (1989).

\bibitem{vglasss}  D. S. Fisher, M. P. A. Fisher, and D. A. Huse, Phys. Rev.
B{\bf \ 43}, 130 (1991).

\bibitem{mcapas}  O. Brunner {\it et al,} Phys. Rev. Lett. {\bf 67}, 1354
(1991); J. -M. Triscone {\it et al, }Phys. Rev. B {\bf 50}, 1229 (1994).

\bibitem{il}  M. I. J. Probert and A. I. M. Rae, Phys. Rev. Lett. {\bf 75},
1835 (1995).

\bibitem{exp1}  F. de la Cruz, D. L\'{o}pez, and G. Nieva, Phyl. Mag. B {\bf %
70}, 773 (1994).

\bibitem{exp2}  D. L\'{o}pez, G. Nieva, and F. de la Cruz, Phys. Rev. B {\bf %
50}, 7219 (1994).

\bibitem{bosones}  D. R. Nelson, Phys. Rev. Lett {\bf 60}, 1973 (1988).

\bibitem{feig}  M. V. Feigel 'man, V. B. Geshkenbein, L. B. Ioffe, and A. I.
Larkin, Phys. Rev. B {\bf 48}, 16641 (1993).

\bibitem{stroud}  Sasik and D. Stroud, Phys. Rev. B {\bf 48}, 9938 (1993).

\bibitem{teitel2}  Y.-H. Li and S. Teitel, Phys. Rev. B {\bf 45}, 5718
(1992).

\bibitem{carneiro}  G. Carneiro, Phys. Rev. Lett {\bf 75}, 521 (1995).

\bibitem{teitel1}  Y.-H. Li and S. Teitel, Phys. Rev. B {\bf 49}, 4136
(1994).

\bibitem{jb1}  E. A. Jagla and C. A. Balseiro, Phys. Rev B {\bf 52}, 4494
(1995).

\bibitem{jbtbp}  E. A. Jagla and C. A. Balseiro, to be published

\bibitem{jb2}  E. A. Jagla and C. A. Balseiro, Phys. Rev. B, in press.
(LD5779BR)

\bibitem{mppp}  S. R. Shenoy, J. Phys. C {\bf 18}, 1543 (1985) and 5163
(1985); J. S. Chung, K. H. Lee, and D. Stroud, Phys Rev B {\bf 40}, 6570
(1989); K. K. Mon and S. Teitel, Phys Rev Lett. {\bf 62}, 673 (1989); R.
Mehrotra and S. R. Shenoy, Europhys. Lett. {\bf 9}, 11 (1989).

\bibitem{metodo}  D. Dom\'{\i}nguez, J. V. Jos\'{e}, A. Karma, and C.
Wiecko, Phys. Rev. Lett. {\bf 67}, 2367 (1991).

\bibitem{tcdeh}  S. Teitel and C. Jayaprakash, Phys. Rev. Lett {\bf 51},
1999 (1983).

\bibitem{mingo1}  D. Dom\'{\i}nguez, N. G. Jensen, and A. R. Bishop,
preprint.

\bibitem{huse}  R. E. Hetzel, A. Sudbo, and D. A. Huse, Phys. Rev. Lett {\bf %
69}, 518 (1992).

\bibitem{nota}  In Ref. \cite{jb1} we had obtained rather similar results to
the ones described here without introducing disorder. This is due to the
following: we use open boundary condition in the ab-plane, thus distorting
the ideal (bulk) configuration of vortices near the surface of the sample.
For the sample size of Ref \cite{jb1} ($L_{ab}=8$) this effect seems to be
important. However, we use now a disordered system in order to assure that
we have a disordered phase at low temperatures.

\bibitem{stauf}  D. Stauffer,{\it \ Introduction to Percolation Theory, }
Taylor and Francis, London, (1987).

\bibitem{villa}  J. Villain, J. Phys. (Paris) {\bf 36}, 581 (1975); E.
Fradkin, B. Huberman, and S. Shenker, Phys. Rev. B {\bf 18}, 4789 (1978).

\bibitem{jbu}  E. A. Jagla and C. A. Balseiro, unpublished.

\bibitem{teitelpc}  S. Teitel, private communication.

\bibitem{ejec1}  G. Brice\~{n}o, M. F. Crommie, and A. Zettl, Phys. Rev.
Lett. {\bf 66}, 2164 (1991).

\bibitem{ejec2}  A. Arribere {\it et al,} Phys Rev B {\bf 48}, 7486 (1993).

\bibitem{dppc1}  D. L\'{o}pez and F. de la Cruz, private communication.

\bibitem{ust}  Strictly speaking, a $1/T_p$ factor remains in the expression
for $\rho _{ab}(\rho _c=0)$. As $T_p$ depends on field, $\rho _{ab}(\rho
_c=0)$ also depends on field. However, this dependence is very tiny, and
would be hard to be seen in the experiments.
\end{references}
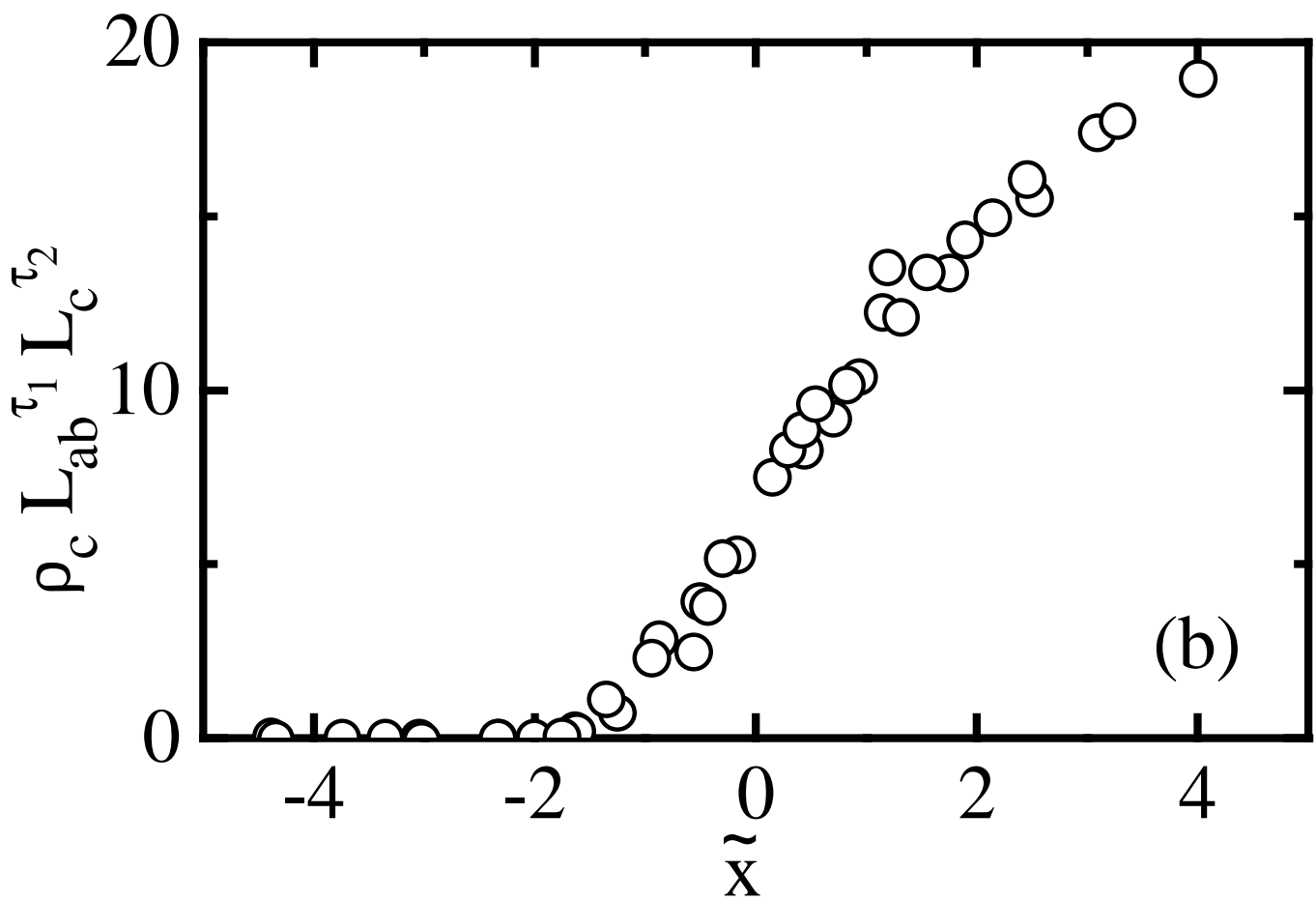
\end{document}